\documentclass[aps,prb,twocolumn,showpacs,superscriptaddress,bibnotes]{revtex4}

\usepackage{graphicx}
\usepackage{dcolumn}
\usepackage{bm}

\begin{document}

\preprint{Submitted to PRB rapid commun.}

\title{Technique for bulk Fermiology by photoemission 
applied to layered ruthenates }

\author{A. Sekiyama}
\author{S. Kasai}
\author{M. Tsunekawa}
\author{Y. Ishida}
\affiliation{Department of Material Physics, Graduate School of 
Engineering Science, Osaka University, Toyonaka, Osaka 560-8531, Japan}%
\author{M. Sing}
\affiliation{Department of Material Physics, Graduate School of 
Engineering Science, Osaka University, Toyonaka, Osaka 560-8531, Japan}%
\affiliation{Experimentalphysik II, Universit{\" a}t Augsburg, 
D-86135 Augsburg, Germany}%
\author{A. Irizawa}
\author{A. Yamasaki}
\author{S. Imada}
\affiliation{Department of Material Physics, Graduate School of 
Engineering Science, Osaka University, Toyonaka, Osaka 560-8531, Japan}%

\author{T. Muro}
\affiliation{Japan Synchrotron Radiation Research Institute, SPring-8, 
Mikazuki, Hyogo 679-5198, Japan}%

\author{Y. Saitoh}
\affiliation{Japan Synchrotron Radiation Research Institute, SPring-8, 
Mikazuki, Hyogo 679-5198, Japan}%
\affiliation{Department of Synchrotron Research, Kansai Research 
Establishment, Japan Atomic Energy Research Institute, SPring-8, 
Mikazuki, Hyogo 679-5148, Japan}%

\author{Y. {\= O}nuki}
\affiliation{Department of Physics, Graduate School of Science, 
Osaka University, Toyonaka, Osaka 560-0043, Japan}%

\author{T. Kimura}
\altaffiliation[Present Address: ]{Los Alamos National Laboratory, 
Los Alamos, NM 87545}
\author{Y. Tokura}
\affiliation{Department of Applied Physics, University of Tokyo, 
Tokyo 113-8656, Japan}

\author{S. Suga}
\affiliation{Department of Material Physics, Graduate School of 
Engineering Science, Osaka University, Toyonaka, Osaka 560-8531, Japan}%

\date{\today}

\begin{abstract}
We report the Fermi surfaces of the 
superconductor Sr$_2$RuO$_4$ and the non-superconductor 
Sr$_{1.8}$Ca$_{0.2}$RuO$_4$ probed by 
bulk-sensitive high-energy angle-resolved photoemission. 
It is found that there is one square-shaped hole-like, 
one square-shaped electron-like and one circle-shaped 
electron-like Fermi surface in both compounds. 
These results provide direct evidence for nesting instability 
giving rise to magnetic fluctuations. 
Our study clarifies that the electron correlation effects are changed 
with composition depending on the individual band. 
\end{abstract}

\maketitle


Clarification of Fermi surfaces (FSs) is fundamental to understand 
the physical properties of functional materials such as 
superconducting transition metal oxides, heavy fermion systems, 
and organic conductors. 
Quantum oscillation measurements by virtue of the de Haas-van Alphen 
or Shubnikov-de Haas effect are known as useful 
techniques to detect bulk FSs. 
However, their electron- or hole-like character and their shape 
cannot be experimentally revealed by these measurements alone. 
In addition, these techniques require low temperatures and almost 
defect-free single crystals, so that they are not easily applicable 
to doped or partially substituted systems such as the high-temperature 
superconductors La$_{2-x}$Sr$_x$CuO$_4$, 
Bi$_2$Sr$_2$CaCu$_2$O$_{8+\delta}$, or the here 
reported Sr$_{1.8}$Ca$_{0.2}$RuO$_4$. 
The number of measurements by using quantum 
oscillations on oxides is actually very few. 
On the other hand, low-energy angle-resolved photoemission (ARPES) 
is known as a tool for probing FSs as well as quasi-particle dispersions 
of correlated electron systems.~\cite{ARPES1} 
However, it is still unclear whether so far reported low-energy ARPES 
($h\nu\lesssim$ 120 eV) results fully reflect bulk 
electronic structures because of its high surface-sensitivity. 
Since high-energy photoemission ($h\nu\ge$ 500 eV) has an advantage 
in probing bulk states,~\cite{ASN,CRSCRG} 
high-energy ARPES with high angular resolution can be a 
complementary and promising technique for the bulk Fermiology of 
solids besides the quantum oscillations measurements. 

It is known that Sr$_2$RuO$_4$ shows "triplet" 
superconductivity~\cite{SRON,SRONMR} , which disappears 
with a very small amount of Ca-substitution.~\cite{NM2000} 
Combination of the quantum oscillation measurements and band-structure 
calculations suggests one hole-like FS sheet centered at $(\pi,\pi)$ 
($\alpha$ sheet) and two electron-like FS sheets centered at (0,0) 
($\beta$ and $\gamma$ sheets) in Sr$_2$RuO$_4$.~\cite{Oguchi,dHvA1,dHvA2} 
On the other hand, so far reported results of low-energy ARPES for 
Sr$_2$RuO$_4$ are controversial 
although ARPES has an advantage in determining the character of FSs. 
Yokoya {\it et al.} have first concluded two 
hole-like and one electron-like FSs.~\cite{Yokoya} 
However, the following ARPES studies~\cite{ARPES1,Puchkov,Dam00} 
have suggested that 
the earlier finding originates from surface states, and 
that the bulk FSs are qualitatively similar to the result of 
the band-structure calculation. 
It has also been reported that a lattice distortion takes place 
at the surface,~\cite{Matz00} giving FSs different from the bulk. 
Thus the characters and shapes of the two-dimensional bulk 
FSs of Sr$_2$RuO$_4$ are experimentally still unclear 
because the reported shapes of the FSs from the low-energy ARPES depend 
on the surface preparation and the excitation photon 
energies.~\cite{Puchkov,Dam00} 
Low-energy ARPES on Sr$_2$RuO$_4$ has shown that the FSs shapes 
measured on the "degraded" surface obtained by cleavage at 180 K 
and fast cooled down seem to be similar to the prediction 
of the band-structure calculation compared with those on the clean 
surface prepared by cleavage at 10 K.~\cite{ARPES1,Dam00} 
In general, photoemission data on cleaner 
surfaces prepared by cleavage at lower temperatures, 
at which surface desorption and diffusion of atoms 
from inside are suppressed, are more reliable. 
Thus the mere similarity of the FSs as obtained by low-energy ARPES 
and theory cannot guarantee that 
the genuine bulk FSs of Sr$_2$RuO$_4$ are really established. 
Besides, the FSs of lightly Ca-substituted 
Sr$_{1.8}$Ca$_{0.2}$RuO$_4$ are not yet clarified at all. 

Compared with low-energy ARPES, high-energy ARPES faces several 
experimental difficulties regarding the detection of quasi-particle 
dispersions and FSs. 
High angular resolution is especially required for high-energy 
ARPES since the momentum resolution not only depends on the angular 
resolution but also on the square root of the photoelectron kinetic 
energy ($\sim h\nu$ in the case of the valence-band ARPES). 
Furthermore, a high photon flux is also required for a practical 
high-energy ARPES measurement because photoionization cross sections 
decrease exponentially with $h\nu$.~\cite{Lindau} 
Recent improvements in both synchrotron light sources and electron 
spectrometers allow us to measure high-energy ARPES spectra 
with high angular and energy resolutions 
facilitating bulk Fermiology. 
In this paper, we show the FSs of Sr$_{2-x}$Ca$_{x}$RuO$_4$ 
probed for the first time by means of high-energy ARPES. 

Single crystals of Sr$_{2-x}$Ca$_{x}$RuO$_4$ ($x = 0, 0.2$) were 
used for the measurements. 
The high-energy ARPES measurements at $h\nu$ = 700 eV were performed 
at BL25SU in SPring-8.~\cite{RSI25} 
The base pressure was about 4 x 10$^{-8}$ Pa. 
The (001) clean surface was obtained by cleaving the samples 
{\it in situ} at the measuring temperature of 20 K. 
The photoelectrons within polar angles of about $\pm 6^{\circ}$ 
with respect to the normal of the sample surface were simultaneously 
collected by using a GAMMADATA-SCIENTA SES200 analyzer, thereby 
covering more than a whole Brillouin zone along the direction of the 
analyzer slit. 
The overall energy resolution was set 
to $\sim$120 and $\sim$200 meV for high-resolution measurements 
and Fermi surface-mapping, respectively. 
The angular resolution was $\pm 0.1^{\circ}$ ($\pm 0.15^{\circ}$) 
for the perpendicular (parallel) direction to the analyzer slit, 
which was experimentally confirmed at BL25SU. 
These values correspond to the momentum resolution of 
$\pm$0.024 {\AA}$^{-1}$ ($\pm$0.035 {\AA}$^{-1}$) 
[6 and 9 \% of $\pi /a$, where $a$ is the lattice constant 
of Sr$_2$RuO$_4$, 3.87 \AA (Ref.~\onlinecite{SRON})] at $h\nu$ = 700 eV. 
The surface cleanliness was confirmed by means of the angle-integrated 
photoemission by the absense of additional spectral 
weight on the higher binding energy side of the intrinsic O $1s$ 
contribution, absence of the possible C $1s$ signal, 
and no peak or hump structure at $9-10$ eV from $E_F$ in the spectra. 
We also measured core-level spectra 
with an energy resolution of 200 meV. 

\begin{figure}
\includegraphics[width=6.5cm,clip]{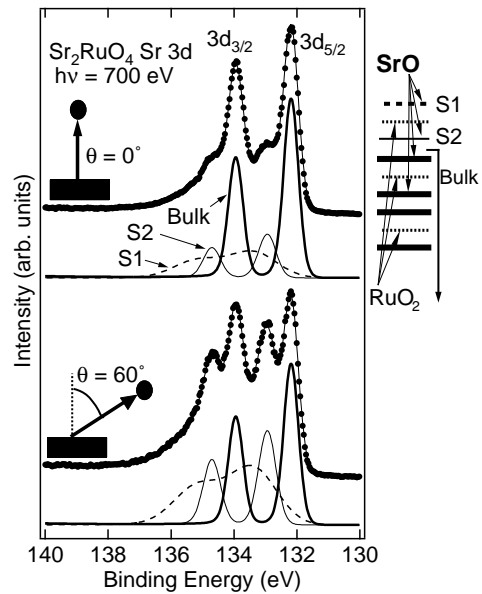}
\caption{\label{Sr3d} Polar-angle ($\theta$) dependence of the Sr 
$3d$ core-level photosmission spectra of Sr$_2$RuO$_4$ (filled circles). 
The spectra are well deconvoluted into three components corresponding 
to the contributions from the top-most SrO surface layer (S1), 
the second SrO surface layer (S2) placed just below the top RuO$_2$ layer, 
and the bulk layers.}
\end{figure}

Figure~\ref{Sr3d} shows the polar-angle ($\theta$) dependence of the 
Sr $3d$ core-level spectra of Sr$_2$RuO$_4$. 
The surface contributions in the spectra is enhenced 
with increasing $\theta$. 
In addition to peaks at 132 and 134 eV originating from the $3d$ 
spin-orbit splitting, two shoulder structures appear on the higher 
binding energy side of the main peaks in the spectrum 
at $\theta$ = 0$^{\circ}$. 
The intensity of these shoulders is remarkably enhanced in the 
more surface-sensitive spectrum at $\theta$ = 60$^{\circ}$. 
Therefore the main peaks and the additional structures are ascribed to 
the bulk and surface components, respectively. 
A more detailed analysis shows that there is actually another 
surface component in the spectra.~\cite{Sr3dAna} 
Due to this analysis, both spectra are well deconvoluted 
into the bulk and two surface contributions (S1 and S2) 
with the intensity ratio of Bulk : S1 : S2 = 0.58 : 0.26 : 0.16 
at $\theta$ = 0$^{\circ}$ and 0.36 : 0.39 : 0.25 at 
$\theta$ = 60$^{\circ}$. 
These ratios are very close to what we expect from the lattice 
constants and the calculated photoelectron mean free path~\cite{Tanuma} 
at the kinetic energy of 565 eV ($\lambda$ = 12.3 {\AA}), namely, 
0.59 : 0.24 : 0.17 at $\theta$ = 0$^{\circ}$ and 0.35 : 0.43 : 0.22 at 
$\theta$ = 60$^{\circ}$. 
Since $\lambda$ at the photoelectron kinetic energy of 700 eV 
is calculated as 13.5 {\AA}, 
the bulk contribution in the valence-band spectra at 
$\theta \sim 0^{\circ}$ is estimated as 63 {\%}. 
It was demonstrated in V$_2$O$_3$ that the surface component is noticeably 
suppressed near the Fermi level ($E_F$) due to its more localized 
character.~\cite{Mo03} 
Namely, the stronger electron correlation in the surface than in the 
bulk leads to less quasi-particle weight near $E_F$ 
and enhanced spectral weight of the surface component away from $E_F$ 
in the spectra of various transition metal oxides. 
Likewise the surface spectral weight is expected 
at higher binding energies away from $E_F$ than the 
bulk spectral weight near $E_F$ in Sr$_{2-x}$Ca$_{x}$RuO$_4$. 

\begin{figure}
\includegraphics[width=8.5cm,clip]{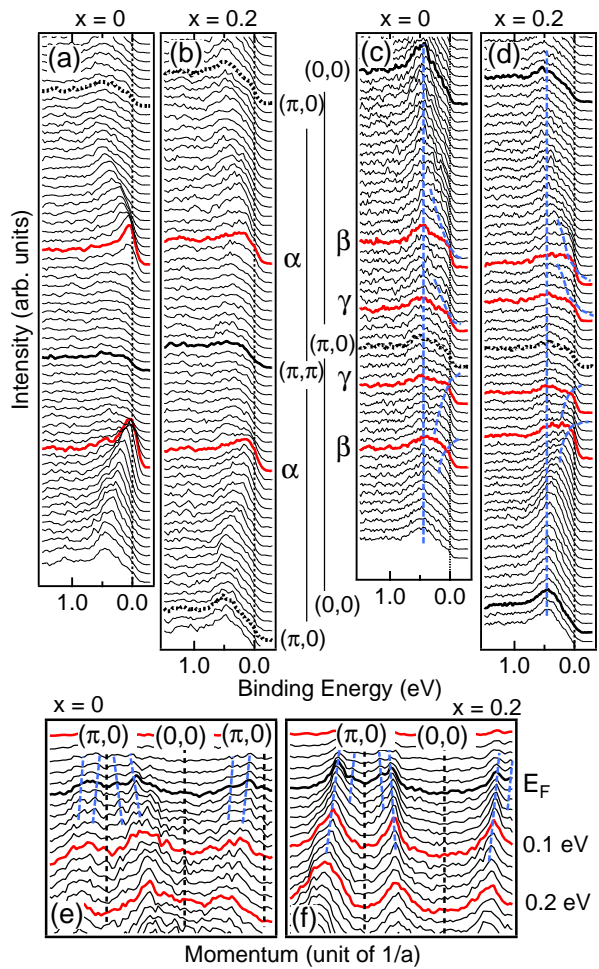}
\caption{\label{EDC} High-energy ARPES spectra near $E_F$ of 
Sr$_{2-x}$Ca$_{x}$RuO$_4$. 
(a) EDCs along the ($\pi$,0)-($\pi$,$\pi$) cut of Sr$_2$RuO$_4$ with 
an energy resolution of $\sim$200 meV. 
The red lines show the spectra where the quasi-particle band crosses 
$E_F$ (i.e., at $k_F$). 
(b) Same as (a) for Sr$_{1.8}$Ca$_{0.2}$RuO$_4$. 
(c) EDCs along the (0,0)-($\pi$,0) direction of Sr$_2$RuO$_4$ 
with an energy resolution of $\sim$120 meV. 
The blue dashed lines indicate the band dispersions. 
(d) Same as (c) for Sr$_{1.8}$Ca$_{0.2}$RuO$_4$. 
(e) MDCs along the (0,0)-($\pi$,0) cut of Sr$_2$RuO$_4$. 
(f) Same as (e) for Sr$_{1.8}$Ca$_{0.2}$RuO$_4$.}
\end{figure}

Let us show the ARPES spectra [energy distribution curves (EDCs)] of 
Sr$_2$RuO$_4$ along the ($\pi$,0)-($\pi$,$\pi$) direction 
in Fig.~\ref{EDC}(a), which demonstrate that a band starts 
from $\sim$0.5 eV at ($\pi$,0), 
approaches to and crosses the Fermi level ($E_F$) 
on going to the ($\pi$,$\pi$) direction. 
The peak width becomes narrower near $E_F$. 
This quasi-particle forms the hole-like FS sheet $\alpha$. 
A similar behavior is observed in Sr$_{1.8}$Ca$_{0.2}$RuO$_4$ as 
shown in Fig.~\ref{EDC}(b). 
However, the peak is broader in the spectra for $x = 0.2$ near the Fermi 
wave vector ($k_F$) compared with those for $x = 0$. 

The ARPES spectra of Sr$_{2-x}$Ca$_{x}$RuO$_4$ 
along the (0,0)-($\pi$,0) cut are shown in Figs.~\ref{EDC}(c) and 
\ref{EDC}(d). 
They are rather complicated because 
there are three quasi-particle bands below $E_F$ in this direction. 
For both compounds, the band forming the $\alpha$ sheet is located 
at $\sim$0.5 eV, 
which shifts hardly between (0,0) and ($\pi$,0), 
while the other two bands forming the $\beta$ and $\gamma$ sheets 
show dispersion and cross $E_F$. 
The behavior of the $E_F$ crossing of the $\beta$ and $\gamma$ 
branches (hereafter abbreviated as $\beta$ and $\gamma$ crossing) 
is also confirmed by the momentum distribution curves (MDCs) 
as shown in Figs.~\ref{EDC}(e) and \ref{EDC}(f), 
and the symmetrized EDCs with respect to $E_F$ 
(not shown in this paper, a similar procedure was used in 
Ref.~\onlinecite{Puchkov}). 
The behavior of the ARPES spectra for Sr$_{1.8}$Ca$_{0.2}$RuO$_4$ is 
qualitatively similar to that of Sr$_2$RuO$_4$ 
whereas subtle differences can be recognized as 
the quasi-particle of the $\beta$ sheet is more prominent for Sr$_{1.8}$Ca$_{0.2}$RuO$_4$ than for Sr$_2$RuO$_4$ 
whereas the $E_F$ crossing of the $\gamma$ sheet is less prominent 
than for $x$ = 0.

\begin{figure}
\includegraphics[width=8.5cm,clip]{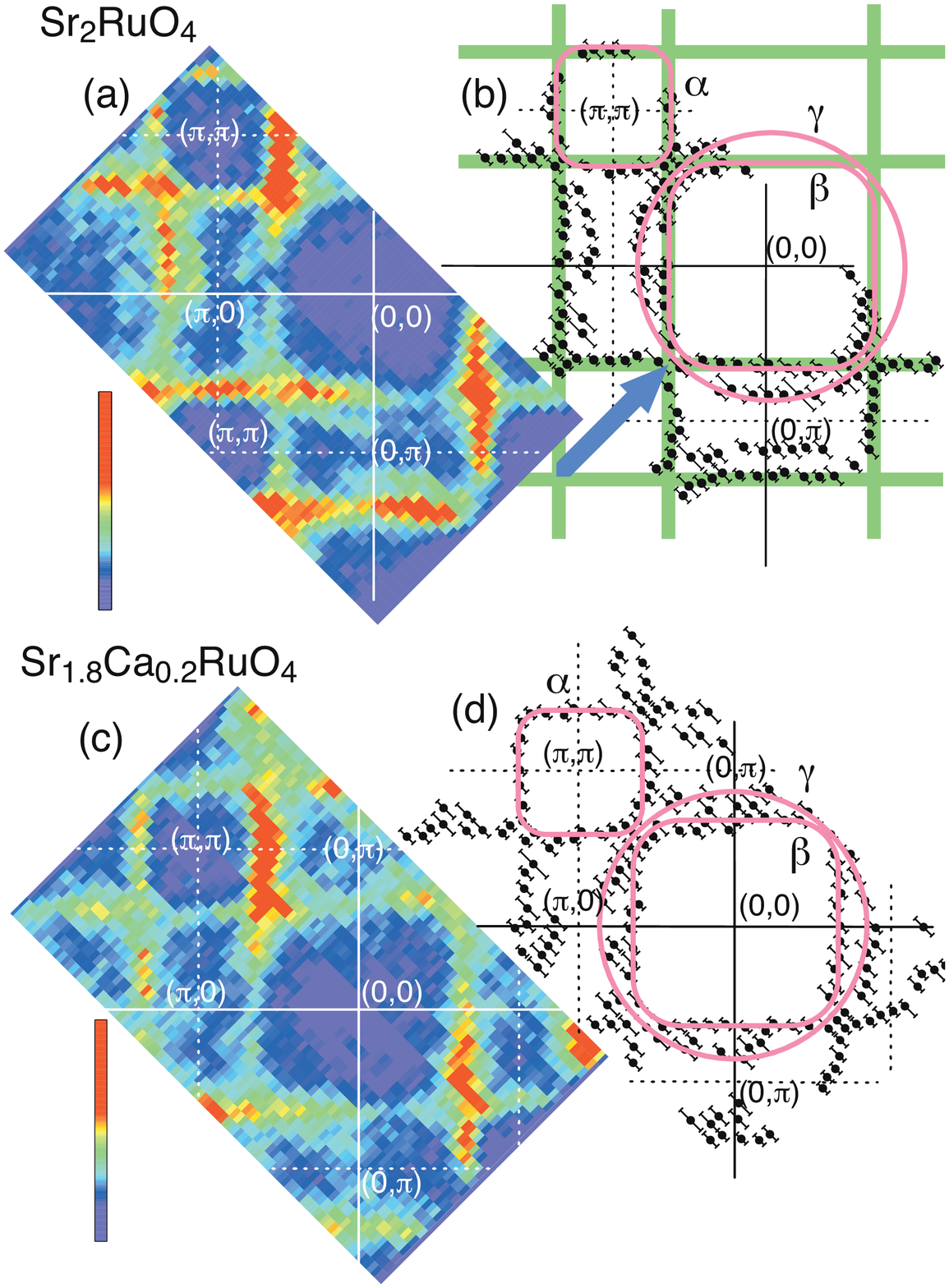}
\caption{\label{Mapp}Fermi surfaces of Sr$_{2-x}$Ca$_{x}$RuO$_4$ 
obtained from the high-energy ARPES.  
(a) Photoemission intensity map integrated from $E_F$ to $-0.1$ eV 
(unoccupied energy side), which represents the FSs 
of Sr$_2$RuO$_4$. 
(b) Estimated $k_F$ and schematically drawn FSs (pink solid curves) 
based on our results. 
The bold green lines show two one-dimensional FSs mutually 
orthogonalized, as a result of the combination of the obtained 
two square-like FSs, $\alpha$ and $\beta$ sheets. 
The blue arrow represents the feasible nesting vector. 
(c) Same as (a) for Sr$_{1.8}$Ca$_{0.2}$RuO$_4$. 
(d) Same as (b) for Sr$_{1.8}$Ca$_{0.2}$RuO$_4$.}
\end{figure}

The photoemission intensity at $E_F$ for Sr$_{2-x}$Ca$_{x}$RuO$_4$ 
and estimated Fermi wave vectors~\cite{criteria} 
$k_F$'s are displayed in Fig.~\ref{Mapp}. 
One can clearly identify the one hole-like ($\alpha$) 
and the two electron-like ($\beta$ and $\gamma$) FS sheets. 
We find that the shapes of the $\alpha$ and $\beta$ sheets 
are square-like while the shape of the $\gamma$ sheet is 
rather circular-like for both compounds. 
These shapes reflect that the $\gamma$ sheet 
is mainly composed of a rather ideally two-dimensional $d_{xy}$ band 
while the other square-shaped sheets are due to the $d_{yz}$ and 
$d_{zx}$ bands, which are to some extent one-dimensional 
in the electronic states. 
The estimated area of each sheet is comparable to the results from the 
quantum oscillations for Sr$_2$RuO$_4$. 
The obtained FSs of Sr$_{1.8}$Ca$_{0.2}$RuO$_4$ are similar to 
those of Sr$_2$RuO$_4$. 
The two-dimensional topology of the FSs 
is also consistent with the prediction from band-structure 
calculations.~\cite{dHvA1,dHvA2} 

The combination of the observed two square-like FS sheets, 
the $\alpha$ and $\beta$ sheets, 
can also be regarded as two one-dimensional FSs 
located at $k_x = \pm Q$ and $k_y = \pm Q$, 
where $Q$ is estimated as $\sim 0.65\pi$ from our high-energy ARPES. 
It has been theoretically predicted 
that FS nesting effect occurs with wave vectors 
${\bf q} = (\pm 2\pi /3,k_y')$, ${\bf q} = (k_x',\pm 2\pi /3)$ 
and especially at ${\bf q} = (\pm 2\pi /3,\pm 2\pi /3)$ 
where $k_y'$ and $k_x'$ are arbitrary.~\cite{Mazin} 
An inelastic neutron scattering study indeed detected 
magnetic fluctuations for ${\bf q}_0 = (\pm 0.6\pi,\pm 0.6\pi)$, 
which could be due to the nesting properties.~\cite{Neu99} 
As shown in Fig.~\ref{Mapp}, the observed FSs give direct evidence 
for the nesting instability 
with ${\bf q} = (\pm 2\pi /3,\pm 2\pi /3)$. 

For Sr$_{1.8}$Ca$_{0.2}$RuO$_4$, the $\gamma$ crossing cannot be 
resolved in some $k$-regions 
whereas the $\beta$ crossing is clearly detected almost everywhere 
in reciprocal space [Figs.~\ref{Mapp}(c) and \ref{Mapp}(d)]. 
This is somewhat different from our FS map of Sr$_2$RuO$_4$, 
where also the $\gamma$-crossing can be resolved almost everywhere 
[Figs.~\ref{Mapp}(a) and \ref{Mapp}(b)] and the spectral weight at $k_F$ 
is comparable between the $\beta$ and $\gamma$ sheets in most of the 
$k$-regions. 
As for the $\alpha$ sheet, a reduced quasi-particle weight at $E_F$ 
is observed for $x = 0.2$ compared with that for $x = 0$ 
as shown in Fig.~\ref{EDC}. 
From these facts we conclude that the electron correlations become 
stronger for the $\alpha$ and $\gamma$ sheets 
by the small amount of Ca-substitution ($x < 0.5$), 
which does not lead any lattice distortion.~\cite{Friedt} 

Although the electron correlation strengths seem to be changed 
depending on the individual band between the superconducting 
Sr$_2$RuO$_4$ and the non-superconducting Sr$_{1.8}$Ca$_{0.2}$RuO$_4$, 
the electronic structures are found to be qualitatively unchanged. 
On the other hand, it has been reported that the 
superconductivity in Sr$_2$RuO$_4$ is easily suppressed 
by impurities and/or defects.~\cite{Mack98,Nishi98} 
From these facts, we can conclude that the superconductivity disappears 
with the Ca-substitution because the substituted Ca ions behave 
as "impurity" and/or induce disorder in Sr$_{2-x}$Ca$_{x}$RuO$_4$, 
and therefore break the superconductivity, 
as proposed by Nakatsuji and Maeno.~\cite{NM2000} 

To conclude, the bulk-sensitive high-energy ARPES study of 
Sr$_{2-x}$Ca$_x$RuO$_4$ has revealed the character and the shape 
of FS sheets, and the nesting instability. 
We are convinced that high-energy ARPES measurements are 
crucial for really revealing the bulk FSs of 
many transition metal oxides. 

We thank T. Satonaka, H. Fujiwara, A. Higashiya, 
P. T. Ernst, A. Shigemoto, and T. Sasabayashi 
for supporting the experiments. 
This work was supported by a Grant-in-Aid for COE Research (10CE2004) 
and Creative Scienrific Research (15GS0213) from 
the Ministry of Education, Culture, Sports, Science and Technology 
(MEXT), Japan. 
M.S. is grateful for financial support by the Japan Society for 
the Promotion of Science (JSPS). 
The ARPES measurements were performed under the approval of the Japan 
Synchrotron Radiation Research Institute (2001A0128-NS-np, 
2002B3009-LS-np, and 2003A4009-LS-np). 


\end{document}